\documentclass[preprint,12pt]{elsarticle}

\usepackage{amsmath,amssymb,amsfonts,bm}
\usepackage{amsthm}
\newtheorem{theorem}{Theorem}

\newtheorem{corollary}{Corollary}

\usepackage{graphicx}
\usepackage{amssymb}

\usepackage{lineno}

\begin{document}

\begin{frontmatter}


\title{On a reduction of the  weighted induced bipartite subgraph problem to the weighted  independent set problem}



\author{Yotaro Takazawa}
\ead{takazawa.y.ab@m.titech.ac.jp}
\author{Shinji Mizuno}
\address{Tokyo Institute of Technology\\ 2-12-1 Ookayama, Meguro-ku, Tokyo 152-8550 Japan}


\begin{abstract}
We study the  weighted induced bipartite subgraph problem  (WIBSP).
The goal of WIBSP is, given a graph and nonnegative weights for the nodes, to find a set $W$ of nodes with the maximum total weight such that a subgraph induced by $W$ is bipartite.
WIBSP is also referred as to the graph bipartization problem or the odd cycle transversal problem.
 In this paper, we show that  WIBSP can  be reduced to  the weighted independent set problem (WISP) where the number of nodes becomes twice and the maximum degree increases by 1. WISP is a well-studied combinatorial optimization problem. Thus,  by using  the reduction and results about WISP,  we can obtain nontrivial approximation and exact algorithms for WIBSP.

\end{abstract}


\end{frontmatter}


\section{Introduction}
Let $G=(V, E)$ be an undirected graph. For $W \subseteq V$,  define $E(W) = \{ \{u, v\} \in E  \ |  \ u,  \ v \in W\} $.
The graph $G(W) = (W,E(W))$ is said to be the subgraph of $G$ induced by $W$. A subset of nodes $W \subseteq V$ is called an independent set of $G$ if $E(W) = \phi $. A graph $G$ is called bipartite if all the nodes can be partitioned into disjoint sets $V_1$ and $V_2$ such that no two nodes within the same set are adjacent. Given a graph $G=(V, E)$ and  nonnegative weights $w_v$ $(v \in V)$, the weighted induced bipartite subgraph problem (WIBSP) is to find  $W\subseteq V$ with the maximum total weight such that the subgraph $G(W)$ of $G$ induced by $W$ is bipartite. WIBSP is also refereed as to  the graph bipartization problem or the odd cycle transversal problem. The weighted independent set problem (WISP) is to find an independent set  $W$ of $G$  with the maximum total weight.

WIBSP has many applications such as VLSI design and the DNA sequencing problem \cite{fouilhoux2012solving}.
For complexity, Choi et al. \cite{choi1989graph} show
that WIBSP is NP-hard even when a graph is planer and the maximum degree is at most 4.
Recently, Ba{\"\i}ou and Barahona \cite{baiou2016maximum} show that
WIBSP can be solved in polynomial time when a graph is planer and the maximum degree is at most 3. 
WIBSP is also studied from multiple viewpoints such as  exponential-time algorithms \cite{raman2007efficient}, approximation algorithms \cite{goemans1998primal,garg1996approximate,klein1990approximation} and fixed parameter algorithms \cite{misra2013parameterized,reed2004finding}.

It is known that WISP can be easily reduced to WIBSP \cite{fouilhoux2012solving}.
In this paper, we show that  WIBSP can also be reduced to WISP where 
the number of nodes  becomes twice  and the maximum degree increases by 1. Since WISP is a well-studied combinatorial optimization problem,  some results about WISP can be used for WIBSP by this reduction. As a result, we can obtain nontrivial  algorithms  for WIBSP.  First, we present a $(\frac{3}{\Delta +3})$-approximation algorithm for WIBSP, where $\Delta$ is the maximum degree of the graph. To our knowledge, this is the first approximation algorithm for WIBSP while there are some approximation algorithms for the minimization version of  WIBSP. Second, we  propose  a $1.439^n  n^{O(1)}$-time algorithm for WIBSP which improves the previous best result of $1.62^n  n^{O(1)}$  by Raman et al. \cite{raman2007efficient}.
Details about these algorithms and related works are shown in Section 3.

\section{Reduction  to the weighted independent set problem}
In this section, we show that WIBSP can be  reduced to WISP.

\begin{theorem}
Given a graph $G=(V,E) \ (V= \{1, \dots, n\})$ and weights $w_v \ (v \in V)$, 
we define the graph $H(G) = (V^1 \cup V^2, E^1 \cup E^2 \cup E^3)$ and weights for the nodes $V^1 \cup V^2$ where
\begin{itemize}
\item $V^1 = \{ v^1_1, \dots, v^1_n\}$
\item $V^2 = \{v^2_1, \dots, v^2_n\}$
\item $E^1 = \{  \{ v^1_i , v^1_j  \} \ (\forall i < j ) \ | \ \{ v_i, v_j \} \in E  \}$
\item $E^2 = \{  \{ v^2_i , v^2_j  \} \  (\forall i < j ) \ | \ \{ v_i, v_j \} \in E  \}$
\item  $E^3 = \{ \{v^1_i, v^2_i\}  \ | \ i \in \{1 \dots n \} \}$
\item  $w_{v_i^1} = w_{v_i^2} =  w_{v_i}$ $(\forall i \in \{ 1 \dots n\} )$.
\end{itemize}
 Then, any independent set $W^I $ of $H(G)$ can be transformed into a set $W^B \subseteq V$  such that $G(W^B)$ is bipartite whose total weight does not change, and vice versa.
\end{theorem}

\begin{figure}[htbp]
  \centering
  \includegraphics[width=8cm]{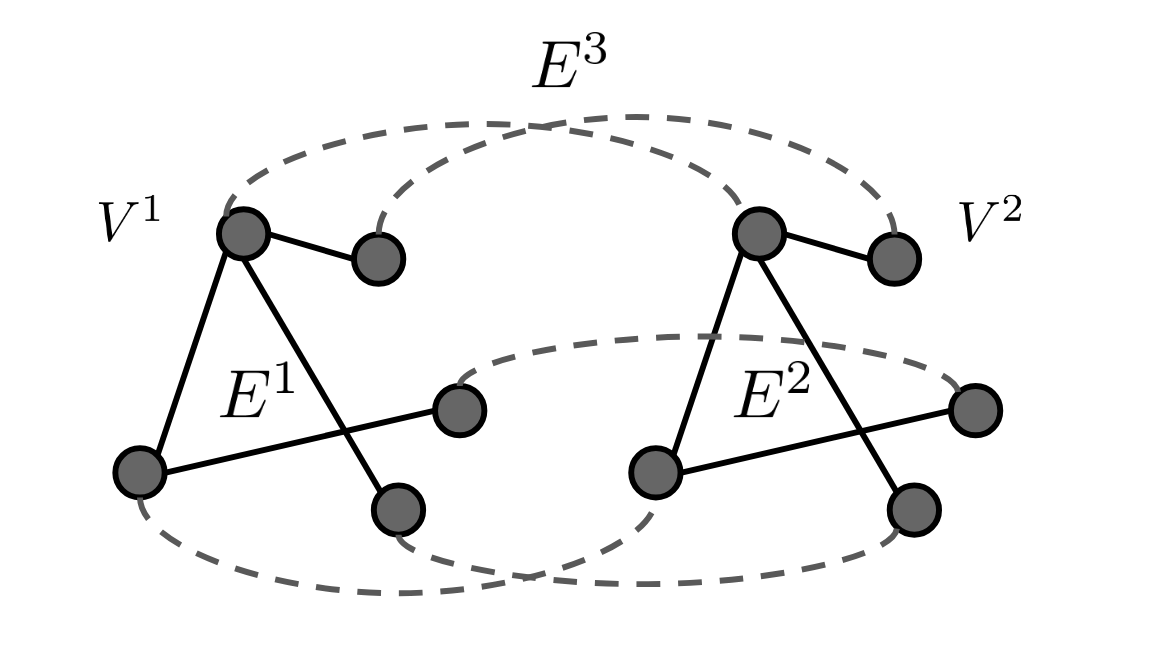}
  \caption{Graph $H(G) = ( V^1 \cup V^2, E^1 \cup E^2 \cup E^3)$}
\end{figure}

\begin{proof}
An example of $H(G)$ is shown in Figure 1.
Let $W^I \subseteq V^1 \cup V^2$ be an independent set of $H(G)$.
Let $W^I_1 =  W^I \cap V^1$ and $W^I_2 =  W^I \cap V^2$.
We define $W^B = W_1^B \cup W_2^B$ where 
$W_1^B = \{ v_i  \in V \ | \ v_i^1 \in W^I_1 \} $ and $W_2^B = \{ v_i  \in V \ | \ v_i^2 \in W^I_2 \} $.
From Figure 2, we can easily see that $W_1^B$ and $W_2^B$ are disjoint and 
 $G(W_1^B)$ and $G(W_2^B)$ are independent sets of $G$.
Therefore,  $G(W^B)$ is bipartite.
We can easily see that the total weight does not change in this transformation.
\begin{figure}[htbp]
  \centering
  \includegraphics[width=8cm]{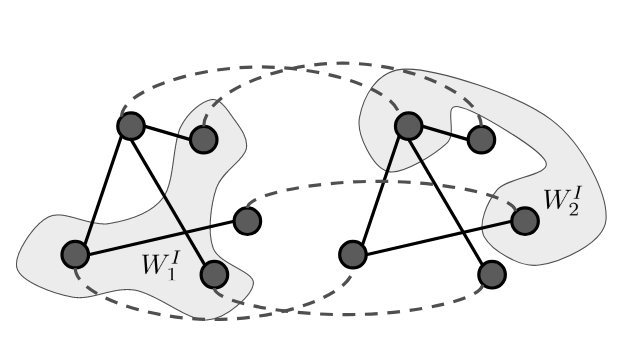}
  \caption{$W_1^I$ and $W_2^I$}
\end{figure}

Next, we will show the converse. 
Let $W^B$ be a subset of $V$  such that $G(W^B)$ is bipartite. Since $G(W^B)$ is bipartite, there exist two  independent sets  $W^B_1$ and $W^B_2$ of $G$, such that $W^B_1 \cup W^B_2 = W^B$ and $W^B_1 \cap W^B_2 = \phi$.
We define $W_1^I = \{ v_i^1  \in V^1 \ | \ v_i \in W^B_1 \} $ and $W_2^I = \{ v_i^2  \in V^2 \ | \ v_i \in W^B_1 \} $.
Then, $W^I = W_1^I \cup W_2^I$ is an independent set of $H(G)$.
The total weight also does not  change.
\end{proof}

Note that  the number of nodes  increase by 2 times and the maximum degree increases by 1 in the reduction of WIBSP to WISP.

\section{Algorithms}
In this section, by using Theorem 1 we obtain approximation and exact algorithms for WIBSP.

\subsection*{Approximation algorithm}
An algorithm is called an $\alpha$-approximation algorithm for a minimization problem (resp., a maximization problem) if it runs in polynomial time and produces a solution whose objective value is less (resp., greater)  than or equal to $\alpha$ times the optimal value.
There are some   approximation algorithms for the minimization version of MWBSP  where we find $W \subseteq V$ with the minimum total weight such that $G(V\backslash W)$ is bipartite.  
Klein et al. \cite{klein1990approximation} present a $O(\log^3 n)$-approximation algorithm.
Garg et al. \cite{garg1996approximate} propose an improved $O(\log n)$-approximation algorithm.
Goemans and Williamson \cite{goemans1998primal} give a $\frac{9}{4}$-approximation algorithm when a graph is planar. To the best of our knowledge, no approximation algorithms are proposed for WIBSP. 
For WISP,  a $\frac{3}{\Delta + 2}$-approximation algorithm is proposed by Halldórsson and Lau \cite{halldorsson2002low}, where $\Delta$ is the maximum degree. By using their algorithm for WISP and Theorem 1, we have the following corollary since the maximum degree increase by 1 in the reduction. Note that for any approximation algorithm for  WISP \cite{kako2009approximation},  we can obtain an approximation algorithm for WIBSP.
%
\begin{corollary}
There is a $(\frac{3}{\Delta +3})$-approximation algorithm for the weighted induced bipartite 
subgraph problem, where $\Delta$ is the maximum degree of the graph.
\end{corollary}

\subsection*{Exact algorithm}

A $1.1996^n n^{O(1)}$-time algorithm for the (unweighted) independent set problem is recently presented by Xiao and Nagamochi \cite{xiao2017exact}, where $n$ is the number of the nodes. By using this algorithm and Theorem 1, we easily get an exact algorithm  for the unweighted induced bipartite subgraph problem, which is a special case of WIBSP where all weights are uniform. In the reduction of Theorem 1, the number of the nodes becomes twice. Thus, by using the algorithm by Xiao and Nagamochi, we have  a $1.1996^{2n}  n^{O(1)} \simeq 1.439^n  n^{O(1)}$ time algorithm which improves the previous best result of $1.62^n  n^{O(1)}$  by Raman et al. \cite{raman2007efficient}.
\begin{corollary}
There is a $1.439^n  n^{O(1)}$-time algorithm for the unweighted induced bipartite subgraph problem,
where $n$ is the number of the nodes.
\end{corollary}

\bibliographystyle{model1-num-names}
\bibliography{sample.bib}







\end{document}